\documentclass[11pt]{article}
 \csname
@addtoreset\endcsname{equation}{section}

\def\bec{\begin{center}}
\def\ec{\end{center}}
\def\a{\alpha} \def\ad{\dot{\a}} \def\hA{{\widehat A}}

\def\b{\beta}  \def\bd{\dot{\b}} 
\def\c{\gamma} \def\cd{\dot{\c}}
\def\C{\Gamma}
\def\d{\delta} 
\def\D{\Delta}
\def\e{\epsilon} 

\def\f{\phi}
\def\F{\Phi}

\def\k{\kappa}
\def\l{\lambda}

\def\m{\mu}
\def\n{\nu}
\def\r{\rho}
\def\s{\sigma}
\def\S{\Sigma}
\def\t{\tau}
\def\th{\theta} 

\def\x{\xi}

\def\pb{{\bar\pi}}

\def\cL{{\cal L}}

\def\cO{{\cal O}}

\def\cN{{\cal N}}

\def\cV{{\cal V}}

\def\hF{\hat{\F}}

\def\yb{{\bar y}}
\def\zb{{\bar z}}

\def\cO{{\cal O}}

\def\ra{\rightarrow}

\let\bm=\bibitem
\def\nn{\nonumber}
\newcommand{\eq}[1]{(\ref{#1})}
\newcommand{\ns}[1]{{\normalsize #1}}

\newcommand{\w}[1]{\\[0.#1cm]}
\def\eqs#1#2{(\ref{#1}-\ref{#2})} \def\det{{\rm det\,}}
\def\be{\begin{equation}}
\def\ee{\end{equation}}
\def\bea{\begin{eqnarray}}
\def\eea{\end{eqnarray}}
\def\ba{\begin{array}}
\def\ea{\end{array}}

\def\ft#1#2{{\textstyle{{\scriptstyle #1}
\over {\scriptstyle #2}}}} \def\fft#1#2{{#1 \over #2}}

\def\scs#1{\section{\bf #1}}
\def\scss#1{\subsection{ #1}}

\begin{document}

\hfill{MIFP-03-09}
\\[-20pt]

\hfill{UU-07-03}
\\[-20pt]

\hfill{hep-th/yymmddd}

\hfill{\today}

\vspace{20pt}
\begin{center}

%%%%%%%%%%%%%%%%%%%%%%%%%%%%%%%%%%%%%%%%%%%%%%%%%%%%%%%%%%%%%%%%%%%%

{\Large\bf Holography in 4D (Super) Higher Spin Theories
\\[10pt] and a Test via Cubic Scalar Couplings}\\

%%%%%%%%%%%%%%%%%%%%%%%%%%%%%%%%%%%%%%%%%%%%%%%%%%%%%%%%%%%%%%%%%%%%

\vspace{30pt}
{\sf\large E. Sezgin}\\[5pt]

{\it\small George P. and Cynthia W.
Mitchell Institute for Fundamental Physics,\\
Texas A\&M University, College Station, TX 77843-4242,
USA}\\[15pt]

{\sf\large P. Sundell}\\[5pt]
{\it\small Department for Theoretical Physics,\\ Uppsala
Universitet, Box 803, SE-751 08 Uppsala, Sweden}

%%%%%%%%%%%%%%%%%%%%%%%%%%%%%%%%%%%%%%%%%%%%%%%%%%%%%%%%%%%%%%%%%%%%

\vspace{30pt} {\bf\Large Abstract}\end{center}

The correspondences proposed previously between higher spin gauge
theories and free singleton field theories were recently extended
into a more complete picture by Klebanov and Polyakov in the case
of the minimal bosonic theory in $D=4$ to include the strongly
coupled fixed point of the 3d $O(N)$ vector model. Here we propose
an $\cN=1$ supersymmetric version of this picture. We also
elaborate on the role of parity in constraining the bulk
interactions, and in distinguishing two minimal bosonic models
obtained as two different consistent truncations of the minimal
$N=1$ model that retain the scalar or the pseudo-scalar field. We
refer to these models as the Type A and Type B models,
respectively, and conjecture that the latter is holographically
dual to the 3d Gross-Neveu model. In the case of the Type A model,
we show the vanishing of the three-scalar amplitude with regular
boundary conditions. This agrees with the $O(N)$ vector model
computation of Petkou, thereby providing a non-trivial test of the
Klebanov-Polyakov conjecture.

\pagebreak

\setcounter{page}{1}  \tableofcontents
\addtocontents{toc}{\protect\setcounter{tocdepth}{2}}

%%%%%%%%%%%%%%%%%%%%%%%%%%%%%%%%%%%%%%%%%%%%%%%%%%%%%%%%%%%%%%%%

\scs{Introduction}

%%%%%%%%%%%%%%%%%%%%%%%%%%%%%%%%%%%%%%%%%%%%%%%%%%%%%%%%%%%%%%%%

A connection between three dimensional free field theory and
massless higher spin (HS) theory in AdS$_4$ was proposed long ago
in \cite{bsst}. Motivated by the advances made in AdS/CFT
correspondence in recent years, this connection was revisited in
\cite{us2} and an increasingly sharpened picture of HS/CFT
correspondence has begun to emerge \cite{h1,h2,h3,h4,holo,kp}. The
basic set up in this correspondence is as follows. Starting from a
large number, $N$ say, of free fields and considering composite
operators invariant under some flavor symmetry group, one
identifies the generating functional with an effective action of a
bulk theory with a perturbation series expansion in powers of
$1/N$ around an AdS vacuum such that couplings of $m$ fields with
$n$ derivatives are of order $L^n N^{-m/2}$ where $L$ is the AdS
radius. This makes sense basically due to the fact that the CFT
correlators factorize in the large $N$ limit. The resulting bulk
theory has two significant properties distinguishing it from the
ordinary gauged supergravities. Firstly, the HS gauge theories
have higher derivative interactions that are not small in units of
the AdS radius, i.e. there is UV/IR mixing, while there is a weak
field expansion scheme corresponding to the $1/N$-expansion
\cite{anal}. Secondly, in addition to the stress-energy tensor the
free field theory has flavor neutral conserved HS currents
implying that the bulk theory has local HS symmetry in addition to
local superdiffeomorphisms.

It is natural to first establish holography in this highly
symmetric phase, and then break HS symmetry by introducing an
additional mass-scale via some less symmetric solution thus
showing holography in less symmetric phases of the theory. The
solution may be of the form of a constant VEV for a dilaton-like
scalar, as has been conjectured in the case of the $\cN=4$ theory
in $D=5$ \cite{us2001,holo}. Here the additional parameter should
be identified with the string tension/coupling $L^4T_s^2=Ng_s$ in
the Type IIB closed string theory on AdS$_5\times S^5$ of radius
$L$ corresponding to finite Yang-Mills coupling $g_{\rm YM}^2=g_s$
on the field theory side. Another possibility is to consider
domain-walls in which scalars with a non-zero mass-term are
running. This has been proposed in the case of the $\cN=8$ theory
in $D=4$ \cite{holo} where the scalar should correspond to the 3d
Yang-Mills coupling. Here the domain-wall is expected to
interpolate between an unbroken phase of M theory on AdS$_4\times
S^7$ and a broken phase described by ordinary supergravity.

In both of the above cases the flavor group is $SU(N)$ and the
unbroken bulk theory contains massless fields as well as massive
fields (some of which are Goldstone modes) corresponding to
bilinear and multi-linear single-trace operators, respectively.
The symmetry breaking solutions discussed above involve massless
as well as massive fields. Another possibility, equivalent to the
proposal by Klebanov and Polyakov \cite{kp}, is to consider domain
wall solutions that involve only massless fields. We shall discuss
this in more detail below after further remarks on the massless
sector of the theory.

The underlying HS symmetry algebra is an infinite extension of the
finite-dimensional AdS (super)group. Arranging the spectrum of
composite operators into multiplets of the HS symmetry algebra
there is a distinguished one formed out of the bilinears. This
multiplet contains the stress-energy tensor, the HS currents as
well as some additional lower spin operators. On the bulk side
this set of operators corresponds to the quasi-adjoint, or twisted
adjoint, representation of the HS algebra which is realized in the
bulk as a zero-form master field that contains the lower spin
fields, the spin $s\geq 1$ curvatures and their derivatives. In
the free boundary field theory, the generating functional of
$SU(N)$ invariant composite operators can be consistently
truncated to the one of the bilinear operators corresponding to
the quasi-adjoint representation \cite{holo}. The corresponding
bulk theory is an interacting and self-contained massless HS gauge
theory.

The massless HS theories have been constructed by Vasiliev in
$D=4$ \cite{v92,v99} and in any $D$ in the case of no
supersymmetry \cite{v03}. All of these theories have a minimal
bosonic truncation consisting of massless fields with spin
$s=0,2,4,\dots$, corresponding to the a quadratic scalar operator
(the mass-term), the stress-energy tensor and the spin
$s=4,6,\dots$ currents of free scalars in $D-1$ dimensions. The HS
gauge theories in $D=4$ with supersymmetry have been further
elaborated upon in \cite{us8,n124,superspace}. The interactions
are constructed by introducing an adjoint one-form master gauge
field and writing the field equations as constraints on the
curvatures of the master fields. In fact, the most compact form of
the constraints makes use of an internal non-commutative
twistor-space and has a remarkably simple form in which all the
curvatures carrying spacetime indices vanish. Hence the local
spacetime dependence can be gauged away completely \cite{v92,v99}
which is the basic source for the UV/IR mixing. It is important to
note, however, that the linearized theory has canonical kinetic
terms, so that it is tachyon and ghost free, while higher
derivatives enter only via interactions.

Turning to the Klebanov-Polyakov conjecture, the generating
functional of bilinear composite operators of the $SU(N)$
invariant $d=3$, $\cN=0$ scalar singleton free field theory is
identical to the generating functional of the 3d $O(N^2-1)$ vector
model expanded around its free UV fixed point. In what follows, we
replace $O(N^2-1)$ by $O(N)$ for notational simplicity. The
generating functional of the free $O(N)$ vector model has been
conjectured to correspond to the minimal bosonic HS gauge theory
in $D=4$ based on the HS algebra $hs(4)$ \cite{holo}. From the
bulk point of view, the precise definition of the generating
functional of Witten diagrams requires fixing the irregular
boundary condition $\D_-=1$ for the bulk scalar \cite{kp}.
However, the other possibility, namely the regular boundary
condition $\D_+=2$ is also available, and it has been proposed
that the resulting generating functional is related to the one for
$\D_-=1$ by a Legendre transformation in the large $N$ limit
\cite{kw}. From the field theory point of view, the Legendre
transformation is realized as double trace deformation
\cite{kw,w1,ber,gk}. Hence, remarkably enough, the regular
boundary condition in the bulk theory corresponds to the strongly
coupled IR fixed point of the $O(N)$ vector model \cite{kp}. The
striking fact that the free and the strongly coupled fixed points
of the $O(N)$ vector model are related simply by a Legendre
transformation has been tested recently by Petkou \cite{pet1}.

The HS/CFT correspondence conjecture also contains an interesting
alternative mechanism for breaking HS gauge symmetries by means of
radiative corrections, in which the Goldstone modes are composite
objects formed out of two massless fields \cite{por}. This
mechanism is available in the case of regular boundary conditions
and the Goldstone modes correspond to the anomalous subleading
$1/\sqrt{N}$-corrections to the HS current conservation laws at
the IR fixed point.

In this paper we elaborate on the role of parity in constraining
the bulk interactions of the 4D HS gauge theories. In particular,
we find two minimal bosonic models obtained as two different
consistent truncations of the minimal $N=1$ model which we refer
to as the Type A and Type B models, respectively. The $\cN=1$
model contains a Wess-Zumino multiplet, and the Type A and B
models retain the scalar or the pseudo-scalar, respectively. We
shall argue that the regular and irregular boundary conditions on
the pseudo-scalar in Type B model yields the generating functional
of 3d Gross-Neveu model expanded around the free and strongly
coupled fixed points in the IR and UV, respectively. We also
propose the holographic duals of the two generating functionals of
the minimal $\cN=1$ supersymmetric HS gauge theory associated with
two distinct boundary conditions on the Wess-Zumino multiplet.
These duals are the free and strongly coupled fixed points of an
$\cN=1$ supersymmetric $O(N)$ vector model.

As a non-trivial test of the holography in the Type A model, we
show the vanishing of the three-scalar amplitude in the case that
the scalar fields obey regular boundary conditions. This agrees
with the result for the $O(N)$ vector model obtained long ago by
Petkou \cite{pet3}.

There exists a generalization of the minimal HS gauge theories in
$D=4$ that involves additional auxiliary fields \cite{v92}. The
introduction of these fields does not change the cubic couplings
of the physical fields. Hence the test of holography presented in
this paper continues to hold in the generalized models as well.

Further aspects of our results and the open problems will be
discussed in the last section. Some of the results of Section 4
overlap with those of \cite{pet2} which appeared during the
preparation of this paper.

%%%%%%%%%%%%%%%%%%%%%%%%%%%%%%%%%%%%%%%%%%%%%%%%%%%%%%%%%%%%%%%%%%%%%%%%%%

\scs{Scalar Couplings in the Minimal $\cN=1$ Model}

%%%%%%%%%%%%%%%%%%%%%%%%%%%%%%%%%%%%%%%%%%%%%%%%%%%%%%%%%%%%%%%%%%%%%%%%%%

In this section we give a brief description of the minimal $\cN=1$
HS model \cite{n124} and show the vanishing of the quadratic
scalar self-couplings in the scalar field equation. We further
demonstrate the vanishing of all higher order scalar contact terms
in the scalar field equation, and discuss briefly the issue of
field redefinitions.

%%%%%%%%%%%%%%%%%%%%%%%%%%%%%%%%%%%%%%%%%%%%%%%%%%%%%%%%%%%%%%%%%%%%%%%%%%

\scss{The Model}\label{sec:model}

%%%%%%%%%%%%%%%%%%%%%%%%%%%%%%%%%%%%%%%%%%%%%%%%%%%%%%%%%%%%%%%%%%%%%%%%%%

{\footnotesize \tabcolsep=1mm
\begin{table}[t]
\bec
\begin{tabular}{|l|cccccccccccl|}\hline
& & & & & & & & & & & & \\
{${(\ell,j)}\backslash s$} & $0$ & \ns{$\ft12$} & $1$ &
\ns{$\ft32$} & $2$ & \ns{$\ft52$} & $3$ & \ns{$\ft72$} & $4$ &
\ns{$\ft92$} & $5$ & $\cdots$
\\
& & & & & & & & & & & & \\
\hline
& & & & & & & & & & & & \\
$(-1,\ft12)$ & $1\!+\!{\bar 1}$ & $1$ & & & & & &
& & & & \\
$(0,0)$ & & & & $1$ & $1$ & & & & &
& & \\
$(0,\ft12)$ & & & & & $1$ & $1$ & & & &
& & \\
$(1,0)$ & & & & & & & & $1$ & $1$ &
& & \\
$(1,\ft12)$ & & & & & & & & & $1$ & $1$
& & \\
\phantom{aa}$\vdots$ & & & & & & & & & & & &
 \\ \hline
\end{tabular}
\ec \caption{{\small  The spectrum of the minimal $\cN=1$ HS gauge
theory arranged into levels labelled by $(\ell,j)$. The entries
represent lowest weight representations $D(E_0,s)$ of $SO(3,2)$
with ground states carrying AdS energy $E_0$ and $SO(3)$ spin $s$.
Each level is an $\cN=1$ multiplet with $s_{\rm max}=2\ell+2+j$.
The AdS energies are given by $E_0=s+1$ for $s\geq 1/2$. The
Wess-Zumino multiplet has two scalar lowest weight states with
$E_0=1$ and $E_0=2$, which must be assigned even and odd parity,
respectively, in the parity invariant $\cN=1$ model.}}
\label{table1}
\end{table}}

The minimal $\cN=1$ HS gauge theory is based on the HS algebra
$hs(1|4)$, whose maximal finite-dimensional subalgebra is
$OSp(1|4)$. The fundamental representation of $hs(1|4)$ is the
$OSp(1|4)$ singleton $D(1/2,0)\oplus D(1,1/2)$. The massless
spectrum of $hs(1|4)$ is the symmetric tensor product of two
$OSp(1|4)$ singletons, and is given in Table \ref{table1}. Once
the parity of the graviton is fixed to be even, the parities of
all the other lowest weight states are fixed by HS symmetry since
a generator carrying $n$ units of AdS energy ($n\in \ft12 \mathbf
Z$) has parity $\exp i\pi n$. It follows that the parity of a
lowest weight state with energy $E_0$ is given by $\exp i\pi
(E_0-1)$. The sign in the parity assignments for the fermions is a
matter of convention.

The basic building blocks of the theory are a master $0$-form
$\widehat{\Phi}$ and a master $1$-form

\be \widehat{A}=dx^\mu \widehat{A}_\mu + dz^\a \widehat{A}_\a+
d\zb^{\ad}\widehat{A}_{\ad}\ , \ee

where $x^\m$ is the spacetime coordinate and $(z^\a,\zb^{\ad})$
are Grassmann even $SL(2,{\mathbf C})$ spinor oscillators. The
master fields are functions of $(x,z,\zb)$ as well as an
additional set of internal oscillators $(y_\a,\yb_{\ad},\x,\eta)$,
where $(y_\a,\yb_{\ad})$ are Grassmann even $SL(2,\mathbf C)$
oscillators and $(\x,\eta)$ are real, Grassmann odd oscillators.
The Grassmann even oscillator algebra is represented by the
following associative $\star$-products\footnote{This formula
corrects a typo in \cite{anal}}:\marginpar{star}

\bea &&\widehat f(z,\zb,y,\yb)*\widehat
g(z,\zb,y,\yb)\nn\\[10pt]&=&\widehat f ~\exp{\footnotesize
\left[i\left(\fft{\overleftarrow{\partial}}{
\partial y_\a} + {\overleftarrow{\partial}\over \partial
z_\a}\right) \left({\overrightarrow{\partial}\over \partial y^\a}
- {{\overrightarrow\partial}\over \partial z^\a}\right)
+i\left({\overleftarrow{\partial}\over \partial {\yb}_{\ad}}
-{\overleftarrow{\partial}\over \partial {\zb}_{\ad}}\right)
\left({\overrightarrow{\partial}\over \partial {\yb}^{\ad}} +
{\overrightarrow{\partial}\over \partial {\zb}^{\ad}}\right)
\right]}~\widehat g\ .\label{star} \eea

The Grassmann odd oscillator algebra is given by

\be \x\star\x=1\ ,\qquad \x\star\eta=-\eta\star\x=\x\eta=-\eta\x\
,\qquad \eta\star\eta=1\ .\ee

At the kinematic level, the oscillator dependence of the master
fields is restricted by the following conditions:

\be {\tau}(\hA)= -\hA\ ,\quad \hA^\dagger =-\hA\ ,\quad\quad
{\tau}(\widehat\Phi)={\bar\pi}(\hF)\ ,\quad
\hF^{\dagger}=\pi(\hF)\star \C\ ,\qquad \C\equiv i\x\eta\
,\label{hf2}\ee

where the $\star$-algebra anti-automorphism $\t$ and automorphisms
$\pi$ and $\bar{\pi}$ are defined by

\bea \t(\widehat f(z,\zb,y,\yb,\x,\eta))&=&\widehat
f(-iz,-i\zb,iy,i\yb,i\x,-i\eta)\
,\\[5pt] \pi(\widehat f (z,\zb,y,\yb,\x,\eta))&=&
\widehat f(-z,\zb,-y,\yb,\x,\eta)\
,\\[5pt]
\pb( \widehat f (z,\zb,y,\yb,\x,\eta))&=&\widehat
f(z,-\zb,y,-\yb,\x,\eta)\ .\eea

The constraints giving rise to the full field equations are
\cite{n124}

\bea \widehat{F}&=& \ft{i}4 dz^\a\wedge dz_\a
\cV(\widehat{\Phi}\star \k\C) + \ft{i}4 d\zb^{\ad}\wedge
d\zb_{\ad} \bar{\cV}(\widehat{\Phi}\star \bar{\k})\ ,\label{c1}\w4
\widehat{D}\widehat{\Phi}&=& 0\ ,\label{c2}\eea

where the curvatures are defined as

\bea \widehat{F}&=& d\widehat{A}+\widehat{A}\star\widehat{A}\ ,
\label{hatF}\w2
\widehat{D}\widehat{\Phi}&=&d\widehat{\Phi}+\widehat{A}\star
\widehat{\Phi}-\widehat{\Phi}\star \bar{\pi}(\widehat{A})\
.\label{DPhi}\eea

and the operators $\kappa$ and ${\bar\kappa}$ as

\be \k=\exp(iy^\a z_\a)\ ,\quad\quad
\bar{\k}=\k^\dagger=\exp(-i\yb^{\ad}\zb_{\ad})\ .\ee

The quantity $\cV(X)$ in \eq{c1} is an expansion in positive
powers of $X$ using the $\star$-product, and
$\bar{\cV}(X)=(\cV(X^\dagger))^\dagger$. Conventionally realized
local Lorentz invariance requires \cite{v99}

\be \cV(X)= b_1X+b_3 X\star X\star X+\cdots\ ,\ee

where $b_1\neq 0$ in order for the theory to have well-defined
linearized field equations. The master constraints \eq{c1} and
\eq{c2} are invariant under the field redefinition $\hF\ra F(\hF)$
where $F$ is a real and odd $\star$-function \cite{us2}. Using
this freedom one can take

\be \cV(X)=X\exp(i\theta(X))\ ,\qquad \theta(X)=
(\theta(X^\dagger))^\dagger=\th_0+\th_2 X\star X+\cdots\
.\label{theta}\ee

The phase-factor $b_1=\exp(i\th_0)$ is inconsequential at the
linearized level, where it can be absorbed into the Weyl tensors.
Its first non-trivial appearance starts at the quadratic level in
the field equations. The parameter $\th_{2n}$ ($n=1,2,\dots$)
appears for the first time in the $(2n+1)$th order in the field
equations.

The parity map $P$ is defined by $P^2=1$ and

\be P(y_\a)=\yb_{\ad}\ ,\qquad P(z_\a)=-\zb_{\ad}\ ,\qquad
P(\x)=\x\ ,\quad P(\eta)=\eta\ ,\ee

from which it follows that $P$ is an automorphism of the
$\star$-algebra, and that $P(\k)=\bar\k$, $P\pi=\pb P$ and
$P\tau=\tau P$. The parity transformations of the master fields
are defined by

\be P(\widehat\Phi)=\widehat\Phi\star\C\ ,\qquad P(\widehat
A)=\widehat A\ .\label{p}\ee

Parity invariance of the master constraints requires $\cV(X)$ to
be real which implies that $\th(X)=0$ in \eq{theta}, that
is\footnote{There are several other equivalent definitions of the
parity map. For example, taking $P(\widehat \Phi)=-\widehat
\Phi\star\C$ implies that the parity invariant interactions are
given by $\cV(X)=iX$.}

\be \cV(X)=X\ .\label{pinvvx}\ee

The parity of the physical fields following from \eq{p} is
discussed in the next section.

The master constraints are integrable, which ensures invariance
under gauge transformations with $x$ and $Z$-dependent parameters.
For the same reason, the spacetime field equations, which follow
from

\be \widehat F_{\m\n}|_{Z=0}=\widehat D_\m\widehat \Phi|_{Z=0}=0\
,\label{stfe}\ee

are invariant under $x$-dependent gauge transformations, which
incorporate spacetime diffeomorphisms as well as the local
supersymmetry transformations. We note that while supersymmetry
takes the standard form at the linearized level, it may be
realized in an unconventional fashion at higher orders due to the
particular nature of the higher derivative corrections to the
constraints and transformation rules.

%%%%%%%%%%%%%%%%%%%%%%%%%%%%%%%%%%%%%%%%%%%%%%%%%%%%%%%%%%%%%%%%%%%%%%%%%%

\scss{The Scalar Field Equation up to Second Order in Weak
Fields}\label{sec:2ndo}

%%%%%%%%%%%%%%%%%%%%%%%%%%%%%%%%%%%%%%%%%%%%%%%%%%%%%%%%%%%%%%%%%%%%%%%%%%

In order to obtain the spacetime field equations from \eq{stfe}
one first uses the components of the master constraints \eq{c1}
and \eq{c2} that carry at least one spinor index to solve for the
$Z$-dependence of $\widehat \Phi$ and $\widehat A$ given an
initial condition

\be A_\m=\widehat A_\m|_{Z=0}\ ,\qquad \Phi=\widehat\Phi|_{Z=0}\
.\ee

These fields have the following expansion in the
$(y,\yb,\x,\eta)$-oscillators:

\bea A_\m&=&\sum_{\ell=0}^\infty \left(A_\m^{(\ell,0)}+
A_\m^{(\ell,1/2)}\right)\ ,\\
\qquad \F&=&\F^{(-1,1/2)}+\sum_{\ell=0}^\infty
\left(\F^{(\ell,0)}+\F^{(\ell,1/2)}\right)\ , \eea

where $A_\m^{(\ell,j)}$ and $\F^{(\ell,j)}$ are given by
($j=0,1/2$)

\bea A_\m^{(\ell,j)}&=&\sum_{m+n+p=4\ell+2+2j}
A^{(\ell,j)}_{\m,p}(m,n)\xi^p\eta^{2j}\ ,\\[10pt]
\F^{(\ell,j)}&=&C^{(\ell,j)}+\pi\Big((C^{(\ell,j)})^\dagger\Big)\star
\C
\ ,\label{aellj}\\[10pt]
C^{(\ell,j)}&=&\sum_{n-m-p=4\ell+2j+3}
\F^{(\ell,j)}_p(m,n)\xi^p\eta^{1-2j}\ .\label{cellj}\eea

Here we use the short-hand notation

\be f(m,n) = \frac1{m!n!}y^{\a_1}\cdots y^{\a_m}\yb^{\ad_1} \cdots
\yb^{\ad_n}f_{\a_1\ldots\a_m\ad_1\ldots\ad_n}\ .\label{not}\ee

The one-form $A_\m$ contains the physical fields of the $\ell\geq
0$ multiplets in the spectrum listed in Table \ref{table1}. The
$(-1,1/2)$ sector of the zero-form $\Phi$ contains the physical
fields of the Wess-Zumino multiplet:

\be C^{(-1,1/2)}= \phi +\yb^{\ad} \bar{\l}_{\ad}\xi+\cdots\
,\label{wz}\ee

where the omitted terms are derivatives of the physical fields.
From \eq{p} it follows that the physical spin $s=2,4,\dots$ fields
have even parity\footnote{The physical spin $s\geq 3/2$ fields
arise in $(e^{-1})_a{}^\m A_\m$, which means that their parity
transformation properties are composed from those of the vierbein
and $A_\m$. Note that $P(x^\mu)=x^\mu$, while $P$ acts
non-trivially on local Lorentz indices. In particular, from \eq{p}
it follows that the anti-symmetric component of the vierbein
$e_{\m\,\a\ad}$ is odd under parity.} and that

\be P(\phi)=\bar \phi\ ,\ee

which means that $\phi$ contains a scalar and a pseudo-scalar:

\be \phi=A+iB\ ,\qquad P(A)=A\,\qquad P(B)=-B\ .\label{ab}\ee

The parity of the fermionic fields is a matter of convention, and
depends on the choice of the phase-factors in the oscillator
expansion \eq{aellj} and \eq{wz}.

The $Z$-dependence of $\widehat \Phi$ and $\widehat A$ can be
obtained in a curvature expansion in powers of $\Phi$. The
resulting form of the constraints \eq{stfe} can be analyzed
further in a modified expansion scheme in which both $\Phi$ and
gauge fields residing in the $s_{\rm max}\geq 5/2$ multiplets
(i.e. the multiplets labelled by $(0,1/2)$, $(1,0)$, $(1,1/2)$,
.... ) are treated as weak fields. The resulting scalar field
equation up to quadratic terms has been obtained for the minimal
bosonic truncation of the $\cN=1$ model in \cite{anal} (see Sect
\ref{sec:trunc}). The scalar field equation can be
straightforwardly generalized to the $\cN=1$ case, and the result
is:

\be (D^\m D_\m+2)\,\phi =\left( D^\mu P^{(2)}_\mu -\ft{i}2
(\s^\mu)^{\a\ad} {\partial\over
\partial y^\a} {\partial\over\partial {\yb}^{\ad}}\,
P^{(2)}_\mu\right)_{Y=\x=\eta=0}\ ,\label{se} \ee

where the supercovariant derivative is defined as

\bea D_\m
\Phi&=&\nabla_\m\Phi+\psi_\m\star\Phi-\Phi\star\bar{\pi}(\psi_\m)\
,\\[5pt] \psi_\m&=&{1\over 2i}(\psi_\m^\a Q_\a-\bar\psi_\m^{\ad}\bar
Q_{\ad})\ ,\qquad Q_\a=y_\a\x\ ,\eea

and

\bea P_\m^{(2)} &=& \Phi\star\bar{\pi}(W_\m)- W_\m\star \Phi
\nn\w2 && +\Bigg(\Phi\star\bar{\pi}({\widehat E}_\m{}^{(1)})
-\widehat{E}_\m^{(1)}\star \Phi+\widehat{\Phi}^{(2)}\star
\bar{\pi}(E_\m)-E_\m\star\widehat{\Phi}^{(2)}\Bigg)_{Z=0}\ ,
\label{p2}\eea

where $W_\m$ contains the HS gauge fields, and we have defined
$E_\m=e_\m+\psi_\m$ and

\bea \widehat{E}^{(1)}_\mu &=&-i e_\m^{\a\ad} \int_0^1 {dt\over t}
\Bigg( \left[{\yb}_{\ad},\hA_\a^{(1)}\right]_*
+\left[\hA_{\ad}^{(1)},y_\a\right]_* \Bigg)_{z\ra tz,\zb \ra
t\zb}+\mbox{$\psi_\m$-terms}\ ,\label{e1}\\[7pt] \widehat\Phi^{(2)}&=&
z^\a\int_0^1dt\left[\Phi\star {\bar\pi}(\hA_\a^{(1)}) -
\hA_\a^{(1)}\star \Phi \right]_{t\rightarrow tz, \zb \rightarrow
t\zb} \nn\\[4pt] &&+~{\zb}^{\ad}\int_0^1 dt \left[\Phi\star
{\pi}({\widehat A}_{\ad}^{(1)}) - \hA_{\ad}^{(1)}\star \Phi
\right]_{t\rightarrow tz, \zb \rightarrow t\zb} \label{phi2}\eea

where

\bea \hA^{(1)}_\a&=& -{i b_1 \over 2}z_\a \int_0^1 tdt
\F(-tz,\yb,\x,\eta) \kappa(tz,y)\star \C\ ,\label{a1}\\[5pt]\hA^{(1)}_{\ad}&=&
-{i \bar{b}_1 \over 2}\zb_{\ad} \int_0^1 tdt
\F(y,t\zb,\x,\eta)\bar\kappa(t\zb,\yb)\ .\label{a1p}\eea

In \eq{a1}, the quantity $\Phi(-tz,\yb,\x,\eta)$ is obtained from
$\Phi(y,\yb,\x,\eta)$ simply by the substitution $y\ra-tz$, and a
similar operation is understood in \eq{a1p}.

%%%%%%%%%%%%%%%%%%%%%%%%%%%%%%%%%%%%%%%%%%%%%%%%%%%%%%%%%%%%%%%%%%%%%%%%%%

\scss{Vanishing of Quadratic Scalar Self-Couplings}\label{sec:3pt}

%%%%%%%%%%%%%%%%%%%%%%%%%%%%%%%%%%%%%%%%%%%%%%%%%%%%%%%%%%%%%%%%%%%%%%%%%%

We are interested in the contribution to $P^{(2)}_\m$ in \eq{p2}
that are quadratic in the physical scalar $\phi$ and its
derivatives. Denoting this contribution by $\widetilde P^{(2)}_\m$
we write

\be P^{(2)}_\m=\widetilde P^{(2)}_\m+\cdots\ .\ee

In computing $\widetilde P^{(2)}_\m$, we first observe that the
linearized Lorentz connection and auxiliary gauge fields in $W_\m$
do not depend on the scalar field. Hence the Lorentz connection
inside the kinetic term in \eq{se} does not contribute to the
quadratic self-couplings, nor does the $W_\m$-terms in
$P^{(2)}_\m$. The remaining terms in $P^{(2)}_\m$ are quadratic in
$\Phi$, which means that we only need to keep track of
contributions to $\Phi$ that are linear in $\phi$ and its
derivatives. These are the components $\Phi^{(-1,1/2)}_0(m,m)$
($m=0,1,2,\dots$) occurring in \eq{cellj}, and which are given by
the $m$th order derivative of $\phi$. The remaining components of
$\Phi$ are either linear in physical fields other than $\phi$, or
quadratic or higher order in physical fields. Thus

\bea \widetilde P^{(2)}_\m&=& \Bigg(\Phi\star\bar{\pi}({\widehat
E}_\m{}^{(1)}) -\widehat{E}_\m^{(1)}\star
\Phi+\widehat{\Phi}^{(2)}\star
\bar{\pi}(E_\m)-E_\m\star\widehat{\Phi}^{(2)}\Bigg)_{
\ba{l}_{Z=0}\\_{\Phi\ra\Phi^{(-1,1/2)}_0}\ea}\ .\label{pf2}\eea

A straightforward manipulation of the first two terms yields

\bea &&\Bigg(\Phi\star\bar{\pi}({\widehat E}_\m{}^{(1)})
-\widehat{E}_\m^{(1)}\star \Phi\Bigg)_{\footnotesize
\ba{l}_{Z=0}\\_{\Phi\ra\Phi^{(-1,1/2)}_0}\ea}\nn\\[10pt]&=&
ie_{\m\,\c\cd}\sum_{m,n=0}^\infty\int_0^1 \int_0^1 dt~dt'~ t^n~
t'^{n+1}
 { n\over
(m!n!)^2}~\Phi^{(-1,1/2)}_{\a(m)\ad(m)}\star
\Phi^{(-1,1/2)}_{\b(n)\bd(n)}\nn\\[10pt]
&\star&\!\!\!\Bigg\{
b_1\e^{\cd\bd_n}\left[y^{\a(m)}\yb^{\ad(m)},\,
z^{\b(n)\c}\yb^{\bd(n-1)}e^{itt'yz}\right]_n\C\nn\\[5pt] &&-\bar b_1
\e^{\c\b_n}\left[\zb^{\bd(n)\cd}y^{\b(n-1)}
e^{-itt'\yb\zb},\,y^{\a(m)}\yb^{\ad(m)}\right]_n\Bigg\}_{Z=0}\
,\label{t1}\eea

where $[A,B]_n\equiv A\star B+(-1)^n B\star A$, we use the
shorthand notation $y^{\a(m)}=y^{\a_1}\cdots y^{\a_m}$, and

\bea \Phi^{(-1,1/2)}_{\a(m)\ad(m)}&=&
\Phi^{(-1,1/2)}_{0\,\a(m)\ad(m)}+(-1)^m\bar
\Phi^{(-1,1/2)}_{0\,\a(m)\ad(m)}\C\ ,\\[5pt]
\Phi^{(-1,1/2)}_{0\,\a(m)\ad(m)}&\sim&
i^m(\s^{\m_1})_{\a_1\ad_1}\cdots
(\s^{\m_m})_{\a_m\ad_m}\Big(\nabla_{(\m_1}\cdots
\nabla_{\m_m)}\f-\mbox{traces}\Big)\ .\eea

Setting $Z=0$ in \eq{t1} enforces $m\geq n+1$, which in turn
implies that at least $m-n+1$ $\yb$-oscillators remains in the
first term of \eq{t1}. It follows that \eq{t1} contains at least
two $y$ or $\yb$-oscillators. Next we consider the second group of
terms in \eq{pf2}, which can be written as

\bea \Bigg(\widehat{\Phi}^{(2)}\star
\bar{\pi}(E_\m)-E_\m\star\widehat{\Phi}^{(2)}\Bigg)_{\footnotesize
\ba{l}_{Z=0}\\_{\Phi\ra\Phi^{(-1,1/2)}_0}\ea}&=&\Bigg({i\over 2}
e_{\m\,\c\cd}\{\widehat\Phi^{(2)},
y^\c\yb^{\cd}\}_\star\Bigg)_{\footnotesize
\ba{l}_{Z=0}\\_{\Phi\ra\Phi^{(-1,1/2)}_0}\ea}\\[5pt] &=&
B_\m+\bar\pi\t(B_\m)+\pi[(B_\m+\bar\pi\t(B_\m))^\dagger]\star\C\ ,
\nn\eea

where

\bea B_\m&=&-{b_1\over
4}e_{\m}^{\c\cd}\sum_{m,n=0}^\infty\int_0^1\int_0^1dt~dt'~(t')^{n+1}
{ 1\over (m!n!)^2}~\Phi^{(-1,1/2)}_{\a(m)\ad(m)}\star
\Phi^{(-1,1/2)}_{\b(n)\bd(n)}\nn\\[10pt]&&\star~
\Big\{z_\d\Big[\Big(y^{\a(m)}\yb^{\ad(m)}\Big)\star
\Big(z^{\b(n)\d}\yb^{\bd(n)}e^{it'yz}\Big)\Big]_{z\ra
tz},\,y_\c\yb_{\cd}\Big\}_{\star}\C_{_{Z=0}}\ . \label{t2}\eea

Setting $Z=0$ forces $y_\c$ to contract $z_\d$. It then follows
that $\yb_{\cd}$ cannot be contracted, since non-vanishing
contributions to $\{A(Y,Z),B(Y,Z)\}_{\star}$ involve a total
number of $YY$ and $ZZ$-contractions that is an even integer. Thus
the number of $\yb$-oscillators in $B_\m$ is at least $|m-n|+1$.
Contracting all the remaining $z$-oscillators inside the
square-bracket requires $m\geq n+1$. Hence there are at least two
$y$ or $\yb$-oscillators in $B_\m$.

We conclude that all contributions to $\tilde P^{(2)}_\m$ contain
at least two $y$ or $\yb$-oscillators. Hence there are no
contributions to the right hand side of the physical scalar field
equation \eq{se} that are quadratic in the physical scalar or its
derivatives.

%%%%%%%%%%%%%%%%%%%%%%%%%%%%%%%%%%%%%%%%%%%%%%%%%%%%%%%%%%%%%%%%%%%%%%%%%%

\scss{Vanishing of All Non-derivative Scalar
Self-Couplings}\label{sec:con}

%%%%%%%%%%%%%%%%%%%%%%%%%%%%%%%%%%%%%%%%%%%%%%%%%%%%%%%%%%%%%%%%%%%%%%%%%%

In this section we strengthen the results of the previous section
to show the vanishing of all non-derivative scalar self-couplings,
i.e. the couplings depending only on the undifferentiated scalar.
To show this it suffices to examine the scalar contact terms in
the full scalar field equation which is given by

\be ( D^\m D_\m+2)\phi=\left(  D^\mu P_\mu -\ft{i}2
(\s^\mu)^{\a\ad} {\partial\over
\partial y^\a} {\partial\over\partial {\yb}^{\ad}}\,
P_\mu\right)_{Y=\x=\eta=0}\ ,\label{sefull} \ee

where

\bea P_\m&=&\Phi\star \bar\pi(W_\m)-W_\m\star \Phi+\nn\\
&&+\sum_{n=2}^\infty\sum_{j=1}^n\left(\widehat\Phi^{(j)}\star\bar\pi(\widehat
E_\m^{(n-j)}+\widehat W_\m^{(n-j)})- (\widehat
E_\m^{(n-j)}+\widehat
W_\m^{(n-j)})\star\widehat\Phi^{(j)}\right)_{Z=0}\ ,\eea

and $\widehat E_\m^{(n)}$ and $\widehat W^{(n)}$ are defined by

\be \widehat E_\m ={1\over 1+\widehat L^{(1)}+\widehat
L^{(2)}+\cdots}~E_\m\ ,\qquad \widehat W_\m={1\over 1+\widehat
L^{(1)}+\widehat L^{(2)}+\cdots}~W_\m\ ,\ee

where

\bea \widehat L^{(n)}(\widehat f)&=& -i\int_0^1{dt\over t}
\Bigg[\widehat{A}^{\a(n)}\star \left({\partial \widehat f\over
\partial z^\a} -{\partial \widehat f\over \partial y^\a}\right)
+\left({\partial \widehat f\over
\partial z^\a} +{\partial \widehat f\over \partial y^\a}\right) \star
\widehat{A}^{\a(n)} \nn\w2
&&\qquad\qquad+\widehat{A}^{\ad(n)}\star \left({\partial \widehat
f\over \partial {\zb}^{\ad}} +{\partial \widehat f\over
\partial {\yb}^{\ad}} \right) +\left({\partial \widehat f\over \partial
{\zb}^{\ad}} -{\partial \widehat f\over \partial
{\yb}^{\ad}}\right) \star \widehat{A}^{\ad(n)}\Bigg]_{z\ra
tz,\zb\ra t\zb} \label{lin}\eea

In what follows we use the notation

\be \widehat f|_\phi\ee

to stand for the scalar contact terms in the quantity $\widehat
f$. In order to obtain the scalar contact terms in \eq{sefull} it
suffices to compute the contributions to $P_\m$ from $\widehat
e_\m|_\phi$. Higher order scalar contributions also enter
\eq{sefull} via the auxiliary gauge fields in $W_\m$, but these
are not of contact type. Recall that $\widehat A_\a$ and $\widehat
\Phi$ are obtained from $\widehat F_{\a\ad}=\widehat
D_\a\widehat\Phi=0$ and $\widehat F^\a{}_\a=i\cV(\widehat
\Phi\star\k)$. We next observe that

\be \Phi|_\phi = \Phi^{(-1,1/2)}_0(0,0)=\f+\bar\phi\C\ .\ee

The first terms in the $\Phi$-expansions of $\widehat A_\a$ and
$\widehat \Phi$ are given in \eq{phi2} and \eq{a1}. From \eq{a1}
it follows that

\be \widehat A_\a^{(1)}|_\f=iz_\a(\f+\bar\phi\C) a^{(1)}(yz)\
,\qquad a^{(1)}(yz)=-b_1/2\int_0^1tdt\exp(ityz)\ .\ee

Using $\bar\pi(\widehat A^{(1)}_\a|_\f)=\widehat A^{(1)}_\a|_\f$
it follows from \eq{phi2} that

\be \widehat \Phi^{(2)}|_\f=0\ .\ee

Iterating this procedure to all orders in the $\Phi$-expansion one
finds that

\be \widehat\Phi|_\f=\f+\bar\phi\C\ ,\qquad \widehat
A_\a|_\f=iz_\a a(\f+\bar\phi\C;yz)\ ,\label{aafull}\ee

where the function $a$ can be solved from the constraint on
$\widehat F_{\a\b}|_\phi$, which reads

\be \partial^\a(z_\a a)+i(z^\a a)\star(z_\a a)={1\over 2}
\cV((\f+\bar\phi\C)\star\kappa\C)\ .\ee

Returning to \eq{lin}, one sees that $\widehat L^{(n)}(e_\m)|_\f$
contains only contributions involving
$e_\m^{\a\ad}[\yb_{\ad},\widehat A^{(n)}_\a]_{\star}$, which
vanish due to \eq{aafull}. Therefore, to all orders we find that

\be \widehat e_\m|_\f=e_\m\ .\ee

Hence, to all orders there is no contribution to $P_\m$ that
depends solely on $e_\m$ and the undifferentiated scalar $\f$. As
a consequence the scalar field equation \eq{sefull} does not
contain any scalar contact interaction terms.

We conclude this section by discussing the role of field
redefinitions in interpreting the results obtained above and in
the previous section. To this end, we write the scalar
contributions to the scalar field equation \eq{sefull} as

\be (\nabla^2+\ft{2}{L^2})~\phi~=\sum_{\tiny\ba{c}k\geq 2,\,n\geq 0\\
p_1+\cdots+p_k=2n\ea}\!\!\!\!\!\!
L^{2n-2}\l^{\{\m_1\dots\m_{p_1}\},\dots,\{\n_1\dots\n_{p_k}\}}(\nabla_{\m_1}\cdot\cdot\nabla_{\m_{p_1}}\phi)
~\cdots~ (\nabla_{\n_1}\cdot\cdot\nabla_{\n_{p_k}}\phi)\
,\label{se2}\ee

where $L$ is the AdS radius and each group of indices is totally
symmetric and traceless. The results of the two last sections
imply that

\be \l^{\{\m_1\dots\m_{p_1}\},\dots,\{\n_1\dots\n_{p_k}\}}=0\
\qquad \mbox{for $k=2$ \ or \ $p_1=\cdots=p_k=0$}\
.\label{excl}\ee

An important property of the HS gauge theory is that the
$\l$-coefficients are fixed numerical coefficients, i.e. they
cannot be taken to be parametrically small. Hence all the higher
derivative terms are of the same order and there is no sense in
which one can take a low energy limit in which these terms become
suppressed. For this reason the absence of the higher order
contact interaction terms is less significant than it would be in
an ordinary effective field theory in which there is a second
length scale, such as the string length, above which the higher
derivative interactions are energetically suppressed and one is
left with only the contact terms.

The coefficients in \eq{se2} can be changed by redefining the
scalar field as

\be\phi~=~\tilde\phi~+\!\!\!\sum_{\tiny\ba{c}k\geq 2,\,n\geq 0\\
p_1+\cdots+p_k=2n\ea}
\!\!\!\!\!\!\t^{\{\m_1\dots\m_{p_1},\dots,\n_1\dots\n_{p_k}\}}
(\nabla_{\m_1}\cdot\cdot\nabla_{\m_{p_1}}\tilde\phi) ~\cdots~
(\nabla_{\n_1}\cdot\cdot\nabla_{\n_{p_k}}\tilde\phi)\ .\ee

In particular, such redefinitions may yield couplings of the types
excluded in \eq{excl}. They do not, however, affect amplitudes,
which is important for the test of holography presented in Section
4.

Given that there are as many $\t$-coefficients as
$\l$-coefficients, it should be possible to eliminate all
$\l$-coefficients with $n\geq 2$ order by order in $n$ for fixed
$k$. The resulting contact terms and the mass-term together yield
the scalar potential, while the $n=1$ coefficients define the
Christoffel symbols on the sigma-model manifold.

In general, the physical field equations obtained from the master
constraints contains arbitrarily high derivatives at any given
non-linear order in the weak field expansion scheme. Whether field
redefinitions can be used to obtain physical equations with only a
finite number of derivatives at any given order in the weak
expansion is not clear and requires further study.

%%%%%%%%%%%%%%%%%%%%%%%%%%%%%%%%%%%%%%%%%%%%%%%%%%%%%%%%%%%%%%%%%%%%%%%%%%

\scs{Minimal Bosonic Type A/B Truncations of The $\cN=1$
Model}\label{sec:trunc}

%%%%%%%%%%%%%%%%%%%%%%%%%%%%%%%%%%%%%%%%%%%%%%%%%%%%%%%%%%%%%%%%%%%%%%%%%%

The minimal $\cN=1$ model admits consistent bosonic truncations in
which we retain $(\widehat \Phi_+,\widehat A_+)$ or $(\widehat
\Phi_-,\widehat A_-)$ defined as

\be \widehat \Phi_{\pm}=i^{^{\ft12(1\pm1)}}\ft12(1\pm
\C)\star\widehat \Phi|_{\xi=\eta=0}\ ,\qquad \widehat
A_{\pm}=\ft12(1\pm \C)\star\widehat A|_{\xi=\eta=0}\
.\label{red}\ee

These master fields obey the $\t$ and reality conditions \eq{hf2}
with $\C$ set equal to $1$ and the parity condition

\be P(\widehat \Phi_\pm)=\pm \widehat \Phi_\pm\ .\ee

Actin with the $\pm$-projections on the $\cN=1$ master constraints
\eq{c1} and \eq{c2} we obtain

\bea \widehat{F}&=& \ft{i}4 dz^\a\wedge dz_\a
\cV_{\pm}(\widehat{\Phi}\star \k) + \ft{i}4 d\zb^{\ad}\wedge
d\zb_{\ad} \bar{\cV}_{\pm}(\widehat{\Phi}\star \bar{\k})\
,\label{c11}\w4 \widehat{D}\widehat{\Phi}&=& 0\ ,\label{c21}\eea

where $\cV_\pm$ are given in terms of the $\cV$-function of the
$\cN=1$ model as

\be \cV_+(X)=\cV(X)\ ,\qquad \cV_-(X)=-\cV(iX)\ .\label{vpm}\ee

We shall refer to the models keeping $(\widehat \Phi_+,\widehat
A_+)$ and $(\widehat \Phi_-,\widehat A_-)$ as the Type A and Type
B models, respectively. The spectrum of the two models are given
by

\bea \mbox{Type A}&:& [D(1/2,0)\otimes D(1/2,0)]_{\rm
s}=D(1,0)\oplus
D(3,2)\oplus D(5,4)\oplus \cdots\ ,\label{specA}\\[5pt]
\mbox{Type B}&:& [D(1,1/2)\otimes D(1,1/2)]_{\rm a}=D(2,0)\oplus
D(3,2)\oplus D(5,4)\oplus \cdots\ ,\label{specB}\eea

where we recall that the parities are given by $(-1)^{E_0-1}$.
From \eq{ab} and $\Phi^{(-1,1/2)}_0(0,0)=\phi+\bar\phi\C$ it
follows that the Type A model retains the scalar
$A=\ft12(\phi+\bar\phi)$, while the Type B model retains the
pseudo-scalar $B=-\ft12 i(\phi-\bar\phi)$.

Starting from a parity invariant $\cN=1$ model, in which case
$\cV(X)$ is given by \eq{pinvvx}, the resulting Type A and B
truncations remain parity invariant, and the corresponding
$\cV_\pm$-functions are given by

\bea \mbox{Type A}&:&\cV_+(X)=X\ ,\\[5pt] \mbox{Type B}&:& \cV_-(X)=-i X\ .\eea

%%%%%%%%%%%%%%%%%%%%%%%%%%%%%%%%%%%%%%%%%%%%%%%%%%%%%%%%%%%%%%%%%%%%%%%%%%

\scs{Holography and a Test via Cubic Scalar
Couplings}\label{sec:hol}

%%%%%%%%%%%%%%%%%%%%%%%%%%%%%%%%%%%%%%%%%%%%%%%%%%%%%%%%%%%%%%%%%%%%%%%%%%

In this section we generalize the Klebanov-Polyakov conjecture
\cite{kp} to the $\cN=1$ model and the Type B model in the case of
parity invariant interactions. We argue that the $\cN=1$ model
admit a family of boundary conditions dual to a line of fixed
points of an $\cN=1$ supersymmetric $O(N)$ vector model, which we
examine in detail at the level of the scalar two-point function.
We also propose that the Type B model is the AdS dual of the 3d
Gross-Neveu model. Finally, the Klebanov-Polyakov conjecture is
verified at the level of cubic scalar amplitudes.

%%%%%%%%%%%%%%%%%%%%%%%%%%%%%%%%%%%%%%%%%%%%%%%%%%%%%%%%%%%%%%%%%%%%%%%%%%

\scss{The $\cN=1$ Model as AdS Dual of Super $O(N)$ Vector
Model}\label{sec:holn1}

%%%%%%%%%%%%%%%%%%%%%%%%%%%%%%%%%%%%%%%%%%%%%%%%%%%%%%%%%%%%%%%%%%%%%%%%%%

In the case of parity invariant interactions the only free
parameter of the minimal $\cN=1$ model is the normalization of the
action\footnote{Strictly speaking, at present only the full field
equations are known. This, however, does not create an obstacle to
the computation of cubic scalar amplitudes. }, which we can take
to be

\be S_{\rm cl}={N\over L^2}\int d^4x \cL\ ,\label{s}\ee

where $N$ is related to the number of free fields in the dual CFT,
and the Lagrangian $\cL$ contains no additional free parameter.
The $1/\sqrt{N}$ expansion of the action, which is obtained by
rescaling the fields by $1/\sqrt{N}$, is equivalent to the weak
field expansion scheme discussed in Section 2. To compute
amplitudes one needs to fix boundary conditions. $OSp(1|4)$
symmetry in itself admits a one-parameter family of boundary
conditions for the Wess-Zumino $(-1,1/2)$-multiplet. To describe
these, we write the boundary behavior of $\phi=A+iB$ as

\be A = r\a_+ + r^2 \b_+ \ ,\qquad B = r\a_- + r^2 \b_-\ ,\ee

and define the $d=3$, $\cN=1$ superfields (our conventions are
given in the the Appendix)\footnote{In what follows we use a
simplified notation in which the dependence on the $\ell\geq 0$
multiplets is suppressed.}:

\bea \Phi_- &=& \a_- + i\bar\th\eta_- +{\bar\th\th\over 2i}\b_+\ ,\\
\Phi_+ &=& \a_+ + i\bar\th\eta_+ +{\bar\th\th\over 2i}\b_-\ ,\eea

where the parities are given by $P(\a_\pm)=\pm\a_\pm$ and
$P(\b_\pm)=\pm\b_\pm$, and the scaling dimensions by
$\D(\a_\pm)=1$, $\D(\b_\pm)=2$ and $\D(\eta_\pm)=3/2$. The
$OSp(1|4)$-invariant boundary conditions are given by

\be B_\l\ :\qquad \Phi_- - \l\Phi_+=J\ ,\label{bl}\ee

where $J$ is a source field and $\l$ is an arbitrary real
parameter which can be taken to be positive without any loss of
generality. The resulting generating functionals for connected
amplitudes are defined by

\be e^{iW_\l[J]}=e^{i S_{\rm cl}[\Phi_\l(J)]}\ ,\ee

where $\Phi_\l(J)$ denotes the solution to the bulk field equation
subject to the $B_\l$ boundary condition. Parity is broken for
$\l\neq 0$, while it is restored at $\l=0$ and $ \l=\infty$, where
the source fields are $\Phi_-$ and $\Phi_+$, respectively.

The $\cN=1$ supersymmetric extension of the arguments given by
Klebanov and Witten in \cite{kw}, suggests that $\Phi_-$ and
$\Phi_+$ are conjugate variables in the $B_0$ theory in the sense
that

\be {\delta W_0\over \delta \Phi_-}=-\Phi_+\ ,\ee

while the opposite should hold in the $B_\infty$ theory. This
strongly suggests that the generating functionals $W_0[\Phi_-]$
and $W_\infty[\Phi_+]$ are related by the following parity and
$OSp(1|4)$ invariant Legendre transformation

\be W_\infty[\Phi_+]=W_0[\Phi_-]+\int d^3x d^2\th \Phi_- \Phi_+\
.\label{n1lt}\ee

The scalar contributions to the second term are given by $\int
d^3x (\b_+\a_+ + \a_-\b_-)$. The relative plus-sign is important,
since $\Phi_\pm$ contains sources for mixed regular and irregular
boundary conditions. The general argument given in \cite{kw},
shows that a scalar field $S$ with kinetic term $-\ft12\e_S \int
d^{d+1}x \sqrt{g}( (\partial S)^2+m^2S^2)$, behaving near the
boundary as $S\sim r^{d-\D}S_0+r^\D \widetilde S_0$, yields
generating functionals $W[S_0]$ and $\widetilde W[\widetilde S_0]$
for correlators of operators $\cO$ and $\widetilde \cO$ of
dimension $\D$ and $\widetilde \D=d-\D$, respectively, related by
$\widetilde W[\widetilde S_0]=W[S_0]-\e_S(2\D-d)\int d^dx
S_0\widetilde S_0$. This leads to the rather puzzling prediction
that $\e_A=-\e_B$. Remarkably enough, this is in perfect agreement
with the result found in \cite{fp} for the bulk-stress tensor,
which we have summarized in \eq{tauab}.

Turning to the holographic description of the $\cN=1$ HS gauge
theory, we begin by observing that the $hs(1|4)$ symmetry requires
the $B_0$ boundary condition. The $hs(1|4)$-invariant spectrum
matches the bilinear operator content of a free $d=3$, $\cN=1$
singleton field theory \cite{holo}. Moreover, rescaling the bulk
fields by $1/\sqrt N$ leads to classical $n$-point amplitudes
proportional to $(1/\sqrt N)^{n-2}$. It is therefore natural to
identify the generating functional of the classical $\cN=1$ HS
gauge theory subject to the $B_0$ boundary condition with that of
the free $\cN=1$ supersymmetric $O(N)$ model \cite{holo}:

\be e^{iW_0}=\left\langle \exp {\sum_{(\ell,j)}\int d^3xd^2\th
\cO_{(\ell,j)}\Phi^{(\ell,j)}_-} \right\rangle_0\ ,\ee

where $\cO_{(\ell,j)}$ and $\Phi^{(\ell,j)}_-$ denote the bilinear
operator and the corresponding source superfield of the level
$(\ell,j)$ supermultiplet. In particular, the scalar operator
coupling to the Wess-Zumino multiplet is given by:

\bea \cO &=& {{c_1}\over{\sqrt N}}W^2 = \cO_1+i\bar\th
\cO_{3/2}+{\bar\th\th\over 2i} \cO_2\nn\\&=& {{c_1}\over{\sqrt
N}}(\varphi\cdot\varphi + 2i \bar \th \psi\cdot \varphi+{
\bar\th\th\over 2i} (i\bar\psi\cdot \psi+2f\cdot \varphi))\ ,
\label{cO}\eea

where $W$ is the off-shell $OSp(1|4)$ singleton superfield in the
$O(N)$ vector representation, with the $\th$-expansion

\be W=\varphi+i\bar\th\psi+{\bar\th\th\over 2i} f\ , \ee

and the constant $c_1=2^{3/2}\pi$ is chosen such that
$\langle\cO\cO\rangle$ has unit strength.

The $\cN=1$ supersymmetric extension of the arguments given by
Witten \cite{w1} suggests that the $B_\l$ boundary condition
corresponds to adding the double-trace deformation $\ft{\l}2\int
\cO^2$ to the free action\footnote{This deformation is the
supersymmetric completion of the $\l\int d^3x \cO_1 \cO_2$
deformation which was argued to be exactly marginal in \cite{w1}.
As we shall discuss in Section \ref{sec:sad}, the $hs(1|4)$ theory
is not invariant under exchange of $A$ and $B$, which means that
there is no invariance under $\l\ra 1/\l$ of the type considered
in \cite{w1}.}:

\be e^{iW_\l}=\left\langle \exp {\sum_{(\ell,j)}\int d^3xd^2\th
\cO_{(\ell,j)}\Phi^{(\ell,j)}_-} \right\rangle_\l\ ,\ee

where $\langle\cdots \rangle_\l$ is evaluated using the action

\bea S_\l&=&\ft{i}4\int d^3x d^2\th D^\a W D_\a W + \ft{\l}2 \int
d^3x d^2\th \cO^2\nn\\ &=& \int d^3x\left(-\ft12 (\partial
\varphi)^2-\ft{i}2\bar\psi\s^\mu\partial_\mu\psi+\ft12f^2+
\ft{\l}2(2\cO_1\cO_2+i\bar\cO_{3/2}\cO_{3/2})\right)\
.\label{sl}\eea

The $\l$-parameter breaks parity as well as HS symmetry, in
agreement with the bulk side of the correspondence. It is known
that \eq{sl} defines a superconformally invariant field theory for
all values of $\l$. The absence of renormalizations of $\l$ can be
demonstrated by examining the short-distance behavior of the
$\exp(-\ft{\l}2 \int d^3x d^2\th \cO^2)$ insertion. The
factorization of correlators in the large $N$ limit implies that

\bea \left\langle \ft12 \left(\ft{i\l}2 \int
\cO^2\right)^2\right\rangle_0&\simeq& -\ft{\l}2 \int dZ dZ' \cO(Z)
G(Z,Z') \cO(Z')\nn\\&=& -\ft{\l}2 \int dZ \cO(Z)\int
d^3x'\ft1{|x'|^2} D^2\cO(Z+Z')\ .\eea

The deformation is well-behaved at short distances, and cannot
produce any infinite nor finite corrections to $\l$, which shows
the vanishing of the $\beta$-function at order $\l^2$. Next, let
us examine the $\l$-dependence of the generating functional by
studying the power-law behavior of the two-point function
$\left\langle \cO(Z)\cO(Z')\right\rangle_\l$. Following Gubser and
Klebanov \cite{gk}, we rewrite the partition function

\be e^{iW_\l[J]} \equiv \left\langle \exp i\int\left(\ft{\l}2
\cO^2+ J\cO \right)\right\rangle_0\ ,\label{pf}\ee

by introducing an auxiliary superfield $\Sigma$, with the
following result

\be e^{iW_\l[J]} = \sqrt{\det\ft{i}{\l}}\int D\S \left\langle\exp
i\int(-\ft{1}{2\l}\S^2+(\S+J)\cO)\right\rangle_0\ .\label{af}\ee

Using the fact that the higher point functions of ${\cal O}$ are
suppressed in the $1/N$ expansion, we approximate

\be  \left\langle \exp i\int(\S+J)\cO\right\rangle_0\simeq
\exp-\ft12
\left\langle\left(\int(\S+J)\cO\right)^2\right\rangle_0\ ,\ee

With the help of this formula, the integral in \eq{af} becomes a
Gaussian and it is readily evaluated to give

\be e^{iW_\l[J]}={1\over \sqrt{\det (1-i\l\widehat
G)}}\exp-{1\over 2} \int dZ dZ' J(Z) \left[{\widehat G\over
1-i\l\widehat G}\right](Z,Z')J(Z')\ ,\label{wg}\ee

where

\be (\widehat G F)(Z)=\int dZ' G(Z,Z')F(Z')\ .\ee

From \eq{wg} it follows that

\be \left\langle \cO(Z)\cO(Z')\right\rangle_\l=\left[{\widehat
G\over 1-i\l\widehat G}\right](Z,Z')\ .\ee

In superspace momentum basis we find

\bea (\widehat G F)(P) &=& \int {dZ\over (2\pi)^{3/2}}~ e^{-iPZ}~
dZ'~ {dP'\over (2\pi)^{3/2}}~ e^{iP(Z-Z')}~ \d^2(\pi')~ G(p')~
{dP''\over (2\pi)^{3/2}}~
F(P'')\nn\qquad\\[5pt]
&=&-(2\pi)^{3/2} p^2 G(p) \widehat\gamma (F(p,\pi))\ ,\eea

where $G(p)$ is given in \eq{gp} and we have defined

\be \widehat\gamma (F(p,\pi))=\int d^2\pi' \exp(-{\bar\pi\s^\mu
p_\mu \pi'\over p^2}) F(p,\pi')\ .\ee

Using

\be \widehat\gamma^2(F(p,\pi))= -{1\over p^2} F(p,\pi)\ee

we find that

\be {\widehat G\over 1-i\l\widehat G}={1\over 1+4\pi^4\l^2
}(\widehat G+i4\pi^4\l)\ ,\ee

from which it follows that

\be \left\langle
\cO(Z)\cO(Z')\right\rangle_\l={1\over{1+4\pi^4\l^2 }}\left\langle
\cO(Z)\cO(Z')\right\rangle_0+\mbox{contact-term}\ .\label{2pt}\ee

Hence, the two-point function has power-law behavior and the
anomalous dimension of $\cO$ vanishes for all values of $\l$ in
the large $N$ limit. The model based on the Lagrangian \eq{sl}
deformed by an additional mass term $i\mu\cO$ has been studied in
\cite{bardeen}. The two-point function \eq{2pt}, including
normalization, is in agreement with the $\m\ra 0$ limit of the
results of \cite{bardeen}. Remarkable, it has been shown in
\cite{bardeen} that for $\mu=0$ the scale invariance can be broken
spontaneously at a particular critical value of the
coupling\footnote{The scale invariance is broken by a dynamically
generated fermion mass $\langle \cO_2\rangle_{\l_{\rm crit}}$. For
fixed $\mu\neq 0$, the two-point function has a pole in momentum
space with mass proportional to $\sqrt{1-(\l/\l_{\rm crit})^2}~\m$
for $\l\sim\l_{\rm crit}$, while for fixed momentum the two-point
function reduces to \eq{2pt} in the limit $\m\ra 0$.} given by
$\l_{\rm crit}=1/4\pi$ (the coupling $\l'$ used in \cite{bardeen}
is given by $\l'=16\pi^2 \l$). It would be interesting to find the
bulk interpretation of this phenomenon.

The above analysis provides evidence for analytical dependence of
$W_\l$ on $\l$. Note that while the deformation does not change
the scaling dimension of $\cO$, it breaks parity invariance of the
theory. Once parity is restored at $\l=\infty$, the parity of the
operator $\cO$ has flipped from its value at $\l=0$. Hence the
scalar components of $\cO$ have the following group theoretic
content:

\bea B_{0}&:&D(1,0)_+ \oplus D(2,0)_-\
,\label{wz+}\\[5pt] B_{\infty}&:& D(1,0)_- \oplus D(2,0)_+\
.\label{wz-}\eea

The continuity in $\l$ also implies that the Legendre
transformation \eq{n1lt} between the generating functionals at
$B_0$ and $B_\infty$ can be reproduced on the CFT side by
integrating out the Lagrange multiplier field $\S$. To see this,
following \cite{gk} we let $\S\ra \S-J$ in \eq{af} and take the
limit $\l\ra \infty$. Up to contact terms and overall
normalization this yields:

\be e^{i W_\infty[\tilde J]}=\int D\S e^{i\int \tilde J \S +
iW_0[\S]}\ ,\ee

where $\tilde J= -J/\l$, which in turn yields the Legendre
transformation \eq{n1lt} upon integrating out $\S$ in the large
$N$ limit.

Finally we propose a few tests of the correspondence at the level
of scalar three-point amplitudes in the case of the $B_0$ and
$B_\infty$ boundary conditions. The $B^3$ and $BA^2$ amplitudes
and the corresponding correlators on the CFT side vanish trivially
due to parity invariance. From the results in Section
\ref{sec:3pt} we know that the remaining cubic $A^3$ and $AB^2$
couplings in the HS gauge theory also vanish. Hence the associated
bulk amplitudes, and the corresponding CFT correlators, vanish
provided the integral over bulk-to-boundary propagators are
finite. Hence, the $AB^2$ amplitude vanishes for $B_0$ boundary
condition, in agreement with the fact that $\langle \cO_1 \cO_2
\cO_2 \rangle_0=0$, while the $A^3$ amplitude vanishes for
$B_\infty$ boundary conditions, which yields the prediction that
$\left\langle \cO_2 \cO_2 \cO_2\right\rangle_\infty=0$. This
prediction is in agreement with the Legendre transformation rule
\eq{amp} \cite{pet1}, and it would be interesting to verify it
directly at strong coupling.

The $A^3$ amplitude subject to the $B_0$ boundary condition and
the $AB^2$ amplitude subject to the $B_\infty$ boundary condition
involve integrals over bulk-to-boundary propagators that require
regularization. In dimensional regularization, for example, both
of these integrals diverge as $(D-4)^{-1}$ in $D$ dimensions. The
$A^3$ amplitude corresponds to the non-vanishing free-field theory
correlator $\left\langle \cO_1\cO_1\cO_1\right\rangle_0$, which
implies that the three-point scalar self-coupling in the bulk
theory is proportional to $D-4$ \cite{pet1}. This in turn implies
the $AB^2$ amplitude is finite as well, which leads to the
prediction that the extremal correlator $\left\langle \cO_2
\cO_1\cO_1 \right\rangle_\infty$ in the strongly coupled theory is
non-vanishing. The above discussion is summarized in Table
\ref{table3}.

{\footnotesize \tabcolsep=1mm
\begin{table}[t]
\bec
\begin{tabular}{|lc|c|c|}\hline& & & \\
& & $A^3$ & $AB^2$  \\
& & &\\
\hline & & & \\
$B_0$  & \quad\phantom{.} Bulk:\quad\phantom{.} & \qquad$g_3 \int
K_1 K_1 K_1$ \qquad\phantom{.}&
\qquad$g_3 \int K_1 K_2 K_2$\qquad\phantom{.} \\& & & \\
 & CFT: & ${1\over x_{12} x_{23} x_{31}}$ & ${1\over x_{12} x_{13}}~\d^3(x_{23})$ \\
& & &\\
\hline & & & \\
$B_\infty$ & Bulk: & $g_3 \int K_2 K_2 K_2$ & $g_3 \int K_2 K_1 K_1$ \\& & & \\
 & CFT: & $\d^3(x_{12})\d^3(x_{23})$ & ${1\over x_{12}^2  x_{31}^2}$\\
& & &  \\ \hline
\end{tabular}
\ec \caption{{\small  Schematic summary of $A^3$ and $AB^2$
amplitudes in AdS$_4$ and their corresponding correlators on the
CFT side. The bulk coupling $g_3$ vanishes in $D=4$ dimensions and
is proportional to $D-4$ in $D$ dimensions. The quantities $K_\D$
($\D=1,2$) are bulk-to-boundary propagators. The integrals $\int
K_2 K_2 K_2$ and $\int K_1 K_2 K_2$ are finite while the integrals
$\int K_2 K_1 K_1$ and $\int K_1 K_1 K_1$ are proportional to
$(D-4)^{-1}$. The correlators are related using the amputation
formula \eq{amp}, and their momentum space representations are
given in the Appendix. }} \label{table3}
\end{table}}

%%%%%%%%%%%%%%%%%%%%%%%%%%%%%%%%%%%%%%%%%%%%%%%%%%%%%%%%%%%%%%%%%%%%%%%%%

\scss{The Type A/B Model as AdS Dual of $O(N)$ Vector/Gross-Neveu
Model}\label{sec:ab}

%%%%%%%%%%%%%%%%%%%%%%%%%%%%%%%%%%%%%%%%%%%%%%%%%%%%%%%%%%%%%%%%%%%%%%%%%%

As explained in Section \ref{sec:trunc}, the $\cN=1$ HS gauge
theory field equations admit two different bosonic truncations,
leading to the minimal bosonic Type A and Type B models which
retain the scalar or the pseudo-scalar of the Wess-Zumino
multiplet, respectively. Taking also the $B_\l$ boundary
conditions defined in \eq{bl} into account, the generating
functional $W_\l[\Phi]$ can be consistently truncated in the cases
of the $B_0$ and $B_\infty$ boundary conditions. The consistent
truncation can also be applied to the Legendre transformation
formula \eq{n1lt}. This yields two pairs of bosonic generating
functionals related by Legendre transformations:

\bea \mbox{Type A}&:&W_\infty[\a_+] = W_0[\b_+]+\int d^3x
\b_+\a_+\ ,
\label{legA}\\
 \mbox{Type B}&:& W_\infty[\b_-]=W_0[\a_-]+\int d^3x \a_-\b_-\ .
 \label{legB}\eea

Turning to their holographic duals, the $hs(4)$ invariant
generating functional $W_0[\b_+]$ corresponds to a free $O(N)$
vector model \cite{holo}. Here we propose that the $hs(4)$
invariant generating functional $W_0[\a_-]$ corresponds to the
free $O(N)$ fermion model. Indeed the spectra of massless fields
of the Type A and B models given in \eqs{specA}{specB} are in
one-to-one correspondence with the bilinear $O(N)$ invariant
operators in the corresponding two free field theories. In the
case of $\cN=1$, the bulk scalars $A$ and $B$ couple to
$\varphi^2$ and $\psi^2$, respectively, which clearly is
consistent with the Type A/B truncations at the level of the
scalar multiplet.

For the higher spin fields the truncation acts slightly
differently. In the spin $s=2$ sector in the $\cN=1$ theory, the
graviton arises from the gauging of the $SO(3,2)$ generators
$M_{AB}$, and a second spin $s=2$ field arises from the gauging of
$\widetilde M_{AB}= M_{AB}\C$. These fields couple to
$T=T(\varphi)+T(\psi)$ and $\widetilde T=T(\varphi)-T(\psi)$,
respectively, where $T(\varphi)$ and $T(\psi)$ are the
stress-energy tensors for the free scalars and fermions,
respectively, and $T$ is the total stress-energy tensor of the
free $\cN=1$ supersymmetric $O(N)$ vector model. From \eq{red} it
follows that the graviton in the Type A and B model arises from
the gauging of $(M_{AB}+\tilde{M}_{AB})/2$ and $(M_{AB}-\widetilde
M_{AB})/2$, respectively, and hence couples to $(T+\widetilde
T)/2=T(\varphi)$ and $(T-\widetilde T)/2=T(\psi)$, respectively.

The Legendre transformations \eq{legA} and \eq{legB} are realized
on the CFT side by adding double-trace deformations to the free
CFTs and taking strong coupling limits \cite{gk,kp} (these
deformations are not truncations of the $\cN=1$ supersymmetric
double-trace deformation used to realize \eq{n1lt} in the $\cN=1$
supersymmetric $O(N)$ vector model). In the case of the Type A
model, this leads to the Klebanov-Polyakov conjecture, according
to which the generating functional $W_\infty[\a_+]$ is the AdS
dual of the strongly coupled IR fixed point of the relevant
$\ft{\l}{2N} \int d^3x (\varphi^2)^2$ deformation of the $O(N)$
vector model.

From the results of Section \ref{sec:3pt} it follows that the
$A^3$ amplitude in the Type A model vanishes in the case of
$D(2,0)_+$ boundary condition. This is in agreement with the
vanishing of the three-point scalar correlator in the strongly
coupled $O(N)$ model \cite{pet3}, and therefore provides a
non-trivial test of the Klebanov-Polyakov conjecture. In the case
of the $D(1,0)_+$ boundary condition, the $A^3$ amplitude needs to
be regularized as discussed in the previous section.

In the case of the Type B model, we conjecture that
$W_\infty[\b_-]$ is the AdS dual of the strongly coupled UV fixed
point of the three-dimensional Gross-Neveu model defined by the
four-fermion interaction $\ft{\l}{2N} \int d^3 (\psi^2)^2$. Though
this irrelevant double-trace deformation is non-renormalizable by
the usual power-counting argument, it is known to be
renormalizable in the $1/N$ expansion, and drives the theory to a
strongly coupled fixed point in the UV where $\D((\psi)^2)=1$.

The vanishing of the $B^3$ coupling in the Type B model and the
corresponding correlator in the Gross-Neveu model follows from
parity invariance, and therefore does not provide a non-trivial
test of the above proposal. A non-trivial test would be the
matching of the graviton-$B^2$ amplitude with the corresponding
CFT correlator. We shall comment on this bulk coupling in Section
\ref{sec:sad} in the context of the perturbative stability of the
HS theory.

%%%%%%%%%%%%%%%%%%%%%%%%%%%%%%%%%%%%%%%%%%%%%%%%%%%%%%%%%%%%%%%%%%%%%

\scs{Summary and Discussion}\label{sec:sad}

%%%%%%%%%%%%%%%%%%%%%%%%%%%%%%%%%%%%%%%%%%%%%%%%%%%%%%%%%%%%%%%%%%%%%

{\footnotesize \tabcolsep=1mm
\begin{table}[t]
\bec
\begin{tabular}{|c|c|c|c|}\hline
&&&\\
Model & Deformation & $\l=0$ & $\l=\infty$ \\
&&&\\
\hline &&&\\
\quad $\cN=1$ \quad\phantom{.}& \quad ${\l\over 2}\int d^2\th d^3x
({\rm tr} W^2)^2$ \quad\phantom{.}&\quad\phantom{.}
$D(1,0)_+\oplus D(2,0)_-$\quad\phantom{.} &\quad\phantom{.}
$D(1,0)_-\oplus D(2,0)_+$\quad\phantom{.} \\
&&&\\
Type A & ${\l\over 2}\int d^3x ({\rm tr} \varphi^2)^2$ &
$D(1,0)_+$ (UV) & $ D(2,0)_+$ (IR) \\
&&&\\
Type B &${\l\over 2}\int d^3x ({\rm tr} \psi^2)^2$ &
$ D(2,0)_-$ (IR) & $D(1,0)_-$ (UV)\\
&&&
 \\ \hline
\end{tabular}
\ec \caption{{\small The field theory deformations and
corresponding boundary conditions at the free fixed point at
$\l=0$ and the strongly coupled fixed point at $\l=\infty$. In the
case of $\cN=1$ the two fixed points are connected by a line of
fixed points. In the case of the Type A and B models the fixed
points are related RG flows.  }} \label{tab:table2}
\end{table}}

We have described a parity invariant minimal $\cN=1$ model and its
bosonic truncations, namely the Type A and B minimal bosonic HS
theories. Both bosonic models have local $hs(4)$ symmetry and
spectrum consisting of massless fields with spin $s=0,2,4,\dots$
each occurring once. The scalar is even under parity in the Type A
model and odd in the Type B model.

In the case of HS invariant boundary conditions the above models
correspond to free field theories on the boundary of AdS
\cite{holo}. The models also admit boundary conditions which break
HS symmetry, and which are related to the HS invariant boundary
conditions by Legendre transformations \cite{kp}. On the field
theory side, the Legendre transformations correspond to strong
coupling limits of various double-trace deformations
\cite{kw,w1,ber,gk}. The various deformations and boundary
conditions are summarized in Table \ref{tab:table2}.

We have examined certain couplings in the minimal $\cN=1$ HS
theory in four dimensions and in particular found that the
quadratic scalar contributions to the scalar field equation
vanish. By truncation, the same holds in the Type A and B models.
In the case of the Type B model, the $B^3$ amplitudes and the
corresponding CFT correlators vanish trivially for both $D(1,0)_-$
and $D(2,0)_-$ boundary conditions by parity invariance. In the
case of the Type A model, the $A^3$ amplitude vanishes for the
$D(2,0)_+$ boundary condition, in agreement with the result for
the strongly coupled $O(N)$ vector model obtained by Petkou
\cite{pet3}. In the case of the $\cN=1$ model, the predictions for
the strongly coupled fixed point based on analyzing cubic scalar
amplitudes are given in Table \ref{table3}.

Another test of the correspondence is to construct domain wall
solutions to the Type A and Type B field equations \eqs{c11}{c21}.
The domain wall should have the topology of AdS spacetime and
break $hs(4)$ down to a possibly infinite dimensional subalgebra
with maximal finite subalgebra $ISO(2,1)$. Moreover, it should
interpolate between an asymptotic AdS region close to the boundary
(UV) and another one in the deep interior (IR), such that $D(1,0)$
scalar fluctuations in the UV interpolate into $D(2,0)$
fluctuations in the IR.

In this paper we have been mainly concerned with the HS/CFT
correspondence in the leading order in the $1/N$-expansion. There
are several issues related to the $1/N$-corrections. To begin
with, on the field theory side the $1/N$-corrections require
$\l\neq 0$. In particular, these corrections show up as anomalies
in the HS current conservation laws. This corresponds to
spontaneous breaking of the HS gauge symmetry in the bulk. In
general, the Goldstone modes in the bulk couple to the anomaly
operators on the field theory side \cite{holo}. In the case of
double-trace deformations, the anomaly of the spin $s$ current has
a double-trace character of a particular form suggesting that the
candidate Goldstone mode for the corresponding massless field is a
composite state formed out of the scalar field and a massless spin
$s-2$ field \cite{por}. In general, Higgsing of a massless spin
$s$ field with parity $(-1)^s$ in the $D(s+1,s)_{(-1)^s}$
representation requires a Goldstone mode in the
$D(s+2,s-1)_{(-1)^{s-1}}$ representation \cite{por}. Stated group
theoretically, $D(s+1+\c,s)$ is irreducible for $\c>0$ while it
decomposes into a massless and a massive irrep in the limit when
the anomalous dimension $\c\ra0$:

\be \lim_{\c\ra0}D(s+1+\c,s)_{(-1)^s}=D(s+1,s)_{(-1)^s}\oplus
D(s+2,s-1)_{(-1)^{s-1}}\ ,\qquad s\geq 1\ .\ee

In the case of the Type A and B models, the Goldstone mode for
spin $s\geq 4$ is contained in $D(2,0)_+\otimes
D(s-1,s-2)_{(-1)^s}$ and $D(1,0)_-\otimes D(s-1,s-2)_{(-1)^s}$, as
follows from

\be D(\D,0)_\e\otimes D(s-1,s-2)_{(-1)^s}=\sum_{j=0}^\infty
\sum_{l=1}^\infty D(\D+s-1+l+j,s-2+j)_{\e(-1)^{s+j+l}}\ ,\qquad
s\geq 4\ .\ee

Moreover, there is no candidate Goldstone mode for the graviton,
as expected. In the case of the $\cN=1$ model, the Goldstone modes
arise in a similar fashion. Thus, in all HS models studied here,
candidate Goldstone modes arise for the boundary conditions
corresponding to the $\l=\infty$ limit on the field theory side.

The fact that the Goldstone modes are composite states suggests
that the actual breaking mechanism involves radiative corrections
to the bulk theory. This raises the question whether the HS gauge
theories provide self-contained and consistent quantum theories of
gravity. This is by no means clear and requires further study.

It is natural to embed the free $O(N)$ vector models into free
singleton matrix models, and consider corresponding AdS duals with
massless as well as massive fields, some of which are candidate
fundamental Goldstone modes \cite{holo}. The free matrix theory
has a more intricate $1/N$ expansion than the free $O(N)$ theory
\cite{h1}, due to mixing between single-trace and multi-trace
operators. However, the corresponding bulk action appears to admit
a consistent truncation to the massless sector, which is
problematic in attempting to interpret it as being an effective
action for a quantum theory. This suggests that loop-corrections
require the $\l=\infty$ boundary conditions associated with the
strongly coupled CFT, though there still remains to resolve the
issue of whether bulk loops are well-defined in the massless
theory with $O(N)$ dual, or if massive fields are required with
dual matrix description. One possibility is that the full quantum
theory requires massive fields, in which case the embedding of the
whole scenario into string theory would be natural
\cite{holo,n2,n3}, and the effective action for the massless
fields can be obtained by integrating out the massive fields once
the HS gauge symmetry has been broken.

Clearly, the stability of the bulk theory at the quantum level is
a highly non-trivial issue, especially given the fact that the
positivity of the total energy is not built into the HS field
equations. In order for the theory to make sense perturbatively it
should be stable when expanded around its AdS vacuum. As a first
step it would be desirable to quadratic contributions to the
stress-energy tensor. Progress in this direction has been made in
\cite{fp} where the quadratic contributions from the scalar field
have been extracted from the full field equations of the Type A
and B models. This result generalizes straightforwardly to the
minimal $\cN=1$ model as follows:

\bea T_{\m\n} &=& \tau_{\mu\nu}(\phi)+\mbox{h.c.}\ ,\\[5pt]
\tau_{\mu\nu}(\phi) &=& - \frac49
g_{\mu\nu}\phi^2+\sum_{k=0}^{\infty} \bigg( a_k g_{\mu\nu}
(\phi_{\r_1\dots\r_{k+1}})^2+b_k\phi_{\m\r_1\dots\r_k}\phi_\n{}^{\r_1\rho_k}
+c_k \phi_{\m\n\r_1\dots\r_k}\phi^{\r_1\dots\r_k} \bigg)\
.\phantom{ergin}\label{tmn}\eea

where
$\phi_{\m_1\dots\m_n}=\nabla_{(\m_1}\cdots\nabla_{\m_n)}\phi-{\rm
traces}$, and $a_k$, $b_k$ and $c_k$ are numerical coefficients
which are given in \cite{fp}. The stress-energy tensors in the
Type A and B models are obtained by substituting $\phi=A+iB$,
which gives

\be \tau_{\mu\nu}(\phi)+\mbox{h.c.}=\tau_{\m\n}(A)-\tau_{\m\n}(B)\
,\label{tauab}\ee

and retaining $A$ or $B$. The relative sign in \eq{tauab} is
surprising and intriguing, both from the point of view of $\cN=1$
supersymmetry and that of stability (though the sign matches the
structure of the supersymmetric Legendre transformation formula as
explained below \eq{n1lt}). By construction, the field equations
are invariant under local $hs(1|4)$ symmetries, including $\cN=1$
supersymmetry, whose explicit form follows from the basic
integrable constraints giving rise to the physical field equations
\cite{anal}. In the view of the above result, it would be
instructive to examine exactly how supersymmetry is realized
on-shell up to the second order in the weak field expansion.

In examining the stability of the scalar fluctuations, one has to
take into account the fact that all terms in $\t_{\m\n}$ are of
the same order, since the coefficients in \eq{tmn} behave as
$2^k/(k!)^2$ for large $k$, while $\nabla^\m\phi_{\m\m_1\dots
\m_k}\sim k^2\phi_{\m_1\dots \m_k}$. It may be possible to make
stability manifest by absorbing all higher derivative terms into a
field redefinition, though the feasibility of such a field
redefinition remains to be seen. On the other hand, it may also be
the case that $\t_{\m\n}$ is positive only for certain boundary
condition on the scalar field (unlike the canonical stress-tensor
which is positive for both boundary conditions). Since
loop-corrections to the HS theory only appear to make sense when
the holographic dual is strongly coupled, a natural outcome of the
stability analysis would be that $(-1)^{\D_{\pm}}\t_{\mu\nu}$ is
positive.

\bigskip

{\bf Acknowledgements}

P.S. is thankful to the String Theory Group at The University of
Roma Tor Vergata and the George P. and Cynthia W. Mitchell
Institute for Fundamental Physics for great hospitality. P.S.
would like to thank M. Berg, M. Bianchi, U. Danielsson, J.
Engquist, F. Kristiansson, R. Leigh, A. Petkou, P. Rajan, A.
Sagnotti and M. Vasiliev and for valuable discussions.

\newpage

%%%%%%%%%%%%%%%%%%%%%%%%%%%%%%%%%%%%%%%%%%%%%%%%%%%%%%%%%%%%%%%%%%%%%

\begin{appendix}

\section{Conventions and Useful Formulae}

In this appendix we summarize our conventions for $\cN=1$, $d=3$
superspace and a few other formula that we use in Section 4.1. We
work in Lorentzian signature, $\eta_{ab}=(-++)$, and with
two-component Majorana spinors\footnote{The computation in Section
\ref{sec:holn1} may equally well be carried out in Euclidean space
using the conventions of \cite{zinn}.}. The Dirac matrices
$(\s^a)_{\a\b}$ are real and symmetric and obey

\be
(\s_a)_\a{}^\b(\s_b)_{\b\c}=\e_{\a\c}\eta_{ab}+\e_{abc}(\s^c)_{\a\c}\
.\ee

We use the north-west-south-east convention,
$\th^\a=\e^{\a\b}\th_\b$ and $\th_\a=\th^\b\e_{\b\a}$, and spinor
bilinears are written as $\bar\th \eta=\th^\a\eta_\a$ and $\bar\th
\s^a\eta= \th^\a(\s^a)_\a{}^\b\eta_\b$. The supercovariant
derivative is defined by

\be D_\a = \partial_\a + i(\s^\mu)_{\a\b}\th^\b\partial_\mu\ ,\ee

The integration over odd superspace coordinates is defined by

\be \int d^2\th \d^2(\th)=1\ ,\ee

where the integration measure and the $\d$-function are defined by

\be \qquad d^2\th ={d\bar\th d\th\over 2i}\ ,\qquad
\d^2(\th)={\bar\th\th\over 2i}\ .\ee

The superspace $\d$-function has the following representation

\be \d^2(\th)\d^3(x)=\int {dP\over (2\pi)^3}e^{iPZ}\ ,\ee

where

\be dP=d^2\pi d^3p\ ,\qquad  PZ= p^a x_a -i\pi^\a \th_\a\ .\ee

Fourier transformation in superspace is defined by

\bea F(Z)&=& \int {dP\over (2\pi)^{3/2}} e^{iPZ} \widetilde F(P)\ ,\\
\widetilde F(P)&=& \int {dZ\over (2\pi)^{3/2}} e^{-iPZ} F(Z)\
.\eea

The operator $\cO$ defined in \eq{cO} has the following two-point
function in the free theory

\be \left\langle \cO(Z)\cO(Z')\right\rangle_0=G(Z,Z')= G(Z-Z')=
{1\over (x-x'+i\bar\th\s\th')^2} ,\ee

where we have defined

\be Z-Z'= (x-x'+i\bar\th\s\th',\th-\th')\ .\ee

The momentum space representation of the two-point function is
given by

\be G(P)=\d^2(\pi)G(p)\ ,\qquad G(p)=\int {d^3x\over (2\pi)^{3/2}}
{e^{-ipx}\over x^2}={(2\pi)^{3/2}\over 4\pi \sqrt{-p^2}}\
.\label{gp}\ee

At the level of three-point functions the Legendre transformation
amounts to amputation, which takes the following form in the case
of scalar operators:

\bea &&\left\langle \phantom{\hat{I}\!\!\!}\cO_{\D_1}(x_1)
\cO_{\D_2}(x_2) \cO_{\D_3}(x_3) \right\rangle_\infty\label{amp1}\\
&=& \int d^3x_1'd^3x_2'd^3x_2' G_{\D_1}^{-1}(x_1,x'_1)
G_{\D_2}^{-1}(x_2,x'_2) G_{\D_3}^{-1}(x_3,x'_3)
\left\langle\phantom{\hat{I}\!\!\!} \cO_{\tilde\D_1}(x'_1)
\cO_{\tilde\D_2}(x'_2) \cO_{\tilde\D_3}(x'_3) \right\rangle_0\
,\nn\eea

where $G_\D(x,y)=\left\langle \cO_\D(x)\cO_\D(y)\right\rangle_0$
and $\tilde\D=3-\D$. The amputation becomes local in momentum
space

\be \left\langle \cO_{\D_1}(p_1) \cO_{\D_2}(p_2) \cO_{\D_3}(p_3)
\right\rangle_\infty =  G_{\D_1}^{-1}(p_1) G_{\D_2}^{-1}(p_2)
G_{\D_3}^{-1}(p_3) \left\langle \cO_{\tilde\D_1}(p_1)
\cO_{\tilde\D_2}(p_2) \cO_{\tilde\D_3}(p_3) \right\rangle_0\
.\label{amp}\ee

The momentum space representation of the non-vanishing three-point
functions of the scalar components of the operator $\cO$ are given
at weak and strong coupling by:

\bea \left\langle
\cO_1(p_1)\cO_1(p_2)\cO_1(p_3)\right\rangle_0&\sim&
{\d^3(p_1+p_2+p_3)\over p_1p_2p_3}\ ,\\
\left\langle \cO_1(p_1)\cO_2(p_2)\cO_2(p_3)\right\rangle_0&\sim&
{\d^3(p_1+p_2+p_3)\over p_1}\ ,\\
\left\langle
\cO_2(p_1)\cO_1(p_2)\cO_1(p_3)\right\rangle_\infty&\sim&
{\d^3(p_1+p_2+p_3)\over p_2 p_3}\ ,\\ \left\langle
\cO_2(p_1)\cO_2(p_2)\cO_2(p_3)\right\rangle_\infty&\sim&
\d^3(p_1+p_2+p_3)\eea

\end{appendix}

\end{document}